\documentclass[aps,prd,twocolumn,amsmath,amssymb,groupedaddress,showpacs]{revtex4-1}
\usepackage{epsf,epsfig,graphics,graphicx}
\usepackage{graphicx}
\usepackage{dcolumn}
\usepackage{bm}
\usepackage{amsmath}
\begin{document}
\title{Teleparallel Poincar\'{e} Cosmology and $\Lambda$CDM Model }
\author{Wenjie Lu}
\author{Wolung Lee}
\email{leewl@phy.ntnu.edu.tw}
\affiliation{Department of Physics, National Taiwan Normal University, Taipei, Taiwan 116, R.O.C.}
\author{Kin-Wang Ng}
\email{nkw@phys.sinica.edu.tw}
\affiliation{Institute of Physics, Academia Sinica, Taipei, Taiwan 115, R.O.C.}

\date{\today}
\begin{abstract}
We apply the teleparallelism condition to the Poincar\'{e} gauge theory of gravity. The resultant teleparallelized cosmology is completely equivalent to the Friedmann cosmology derived from Einstein's general theory of relativity. The torsion is shown to play the role of the cosmological constant driving the cosmic acceleration. We then extend such theory to include the effect of spin and explore the possibility of accounting for the current accelerating universe by a spinning dark energy.
\end{abstract}

\maketitle
\section{Introduction}

Recent observational data from type Ia supernovae, cosmic microwave background (CMB) anisotropies, and large scale structure, concordantly prevail an accelerating flat universe containing a mixture of matter and a preponderant smooth component with effective negative pressure~\cite{Ia1,Ia2,wmap7,sdss}. Though suffering from the so-called cosmological constant problem and the coincidence problem~\cite{cc,coin}, the model of cold dark matter with a cosmological constant in the framework of general relativity (GR), i.e. $\Lambda$CDM model, thus has become the standard scenario in cosmology over the past decade. There exist, however, other options to account for the much perplexing accelerated expansion of the universe, among which the generalized teleparallel gravity, or the $f(T)$ gravity~\cite{ft1,ft2,ft3,ft4,ft5}, has attracted lots of attention recently.

The $f(T)$ theory originates from the teleparallel equivalent of general relativity (TEGR) \cite{TGoverview, introductionTG,tegrev} which is based upon the notion of absolute parallelism (teleparallelism) initiated by Einstein in an unsuccessful attempt of unifying gravitation and electromagnetism~\cite{einstein}. As an alternative geometrical formulation of GR, the TEGR employs a non-trivial dynamical tetrad field, $e_{i}\!^{\mu}$, determined by a given metric to define a linear Weitzenb\"{o}ck connection, $\Gamma^{\mu}\!_{\alpha \beta}=e_i\!^\mu\partial_\beta e_\alpha\!^i$. This particular choice of Weitzenb\"{o}ck connection makes the curvature tensor vanish. Thus, the spacetime is flat and the gravitational degrees of freedom are completely specified by the Weiztenb\"{o}ck torsion, ${T}^{\mu}\!_{\alpha \beta}=\Gamma^{\mu}\!_{\alpha \beta}-\Gamma^{\mu}\!_{\beta \alpha}$. This equivalence between teleparallel gravity and GR implies that both torsion and curvature are capable of offering effectively the same picture but with different interpretations regarding the gravitational field. For example, particle geodesics in GR are now replaced by trajectories depicted by a force equation similar to that of Lorentz force in electrodynamics~\cite{torsionforce}. Furthermore, it is argued that the principle of general covariance ultimately prefers torsion to curvature~\cite{TGreapprisal}. 

Generalizations of teleparallel gravity basically supersede the Lagragian density $T$ of the TEGR with various algebraic functions of the torsion scalar, $f(T)$. The torsion tensor so defined contains only products of first derivatives of the tetrads and consequently gives rise to second order field equations. Evidently, for the sake of computation, this feature is regarded as a significant advantage of the $f(T)$ theory when comparing to other modified gravity theories, such as the $f(R)$ theory~\cite{frrev}.

However, these general teleparallel gravity theories are certainly not free of pathologies: they do not respect local Lorentz covariance~\cite{FFinf,barrow1}, i.e. a local Lorentz transformation would inevitably spoil the absolute parallelism except in the simplest TEGR case. It has been shown~\cite{barrow2} that the local Lorentz invariance cannot even be regained by adding a spin connection back to the action. Moreover, since a proper tetrad field used to construct a teleparallel structure cannot be uniquely specified by a given metric, the lack of local Lorentz symmetry will cause difficulties in formulating a satisfactory cosmological model to account for the late time accelerated expansion~\cite{FFcos}. In light of the fact that the Weitzenbo\"{o}ck connection is simply a special choice of the affine structure to make the curvature tensor identically null, to avoid those defects of the $f(T)$ gravity while retaining the notion of absolute parallelism, we thus resort to the more general gravitational theory: the Poincar\'{e} gauge theory (PGT) of gravity~\cite{PGTHehl,BnH,3lecturesPGT}.    

The PGT includes two independent local translational and rotational potentials, the tetrad and affine connection, which respectively correspond to the torsion and curvature in a Riemann-Cartan geometry. The gauge structure and geometric properties of Riemann-Cartan geometry allow the PGT to become an alternative gravity theory to GR. In particular, when the torsion vanishes, the PGT reduces to GR and we recover a Riemann geometry. In the other hand, the PGT would reduce to a generalized teleparallel theory of gravity by imposing the condition of absolute parallelism. These features provide us a sound framework to inspect the dynamics induced by the teleparallelism and compare the result with that from the standard GR.

Prior to the discovery of the accelerating universe, a few PGT cosmological models have been worked out in detailed by Goenner and M\"{u}ller-Hoissen~\cite{goenner}. More recently, attempts have been made to interpret the source driving the accelerated expansion by the PGT~\citep{OscillatingUniverse,SNY,lxz2012,gengpgt}. 

In our study here, we explore the possibility for explaining the cosmic acceleration by the teleparallelized PGT (TPGT). In such a theory, the form of torsion is strongly constrained by the teleparallelism condition. We will show that, by adopting the constrained form of torsion, the resulted formulation governing a homogeneous and isotropic universe is identical to that obtained from GR. Therefore, one asserts that the TPGT is simply a generalized TEGR in essence. As a consequence, the dynamical role played by the cosmological constant along the cosmic evolution and its connection to the Poincar\'{e} torsion will be unraveled. 

Moreover, we investigate the possibility of a spinning fluid as the source driving the cosmic acceleration by including the spin effect in the TPGT. Unlike the Einstein-Carton theory~\cite{GRspin,spinfluid3} which is considered as a natural extension of GR, the gravitational effect in TPGT is characterized by torsion, not by curvature. Thus, the usual formulation of the Weyssenhoff fluid~\cite{spinfluid1,spinfluid2} does not really apply in our case. In particular, it has been shown that the dust Weyssenhoff fluid model is unable to serve as an alternative to dark energy~\cite{spinfluid4}. However, if the spin tensor of the cosmic fluid takes a specific form relating to the Poincar\'{e} torsion scalar and the spin density, we are able to show that the formulation describing the Poincar\'{e} cosmology is actually equivalent to that of the standard Friedmann cosmology plus a spinning dark energy. Under certain circumstances, the cosmological constant can be regarded as a sort of spinning vacuum energy yet to be scrutinized.   

This paper is organized as follows: In Sec. II we summarize the essential and necessary ingredients of the PGT to set the framework for our investigation. Section III explores the form of the Poincar\'{e} torsion by imposing the teleparallelism constraint which requires that all components of the curvature tensor vanish identically. We then explore the Poincar\'{e} cosmology in Sec. IV and reveal the connections between the cosmological constant and the torsion.  Finally, we summarize our findings and discuss the implications in Sec. V.

%
%
\section{The Essential Poincar\'{e} gauge field theory of gravity}
\subsection{The torsion and the curvature}
The Poincar\'{e} gauge field theory of gravity employs two independent local translation and rotation gauge potentials, the orthonormal frame field (tetrad) $e_{i}\!^{\alpha}$ and the Lorentz connection $\Gamma_{i\beta}\!^{\alpha}$, to characterize both the curvature and torsion in a Riemann-Cartan spacetime.  By means of the covariant derivative $D_i$, the two field strength tensors associated with the gravitational potentials are the torsion  
\begin{equation} 
F_{ij} \! ^{\alpha } \equiv 2D_{[i} e_{j]} \! ^{\alpha } =2\left(\partial _{[i} e_{j]} \! ^{\alpha } +\Gamma _{[i|\beta } \! ^{\alpha } e_{|j]} \! ^{\beta } \right) ,
\end{equation}
and the curvature
\begin{equation}
R_{ij\alpha } \! ^{\beta } \equiv 2D_{[i} \Gamma _{j]\alpha } \! ^{\beta } =2\left(\partial _{[i} \Gamma _{j]\alpha } \! ^{\beta } +\Gamma _{[i|\gamma } \! ^{\beta } \Gamma _{|j]\alpha } \! ^{\gamma } \right) .
\end{equation}
Here the Latin indices $i,j,...$ denote the holonomic (coordinate) base, and the Greek letters ${\alpha},{\beta},...$ represent the anholonomic (Lorentz) indices. The reciprocal frame $e^{i}\!_{\mu}$ satisfies $e^{i}\!_{\mu}e_{i}\!^{\nu}={\delta}_{\mu}\!^{\nu}$ and $e^{i}\!_{\mu}e_{j}\!^{\mu}={\delta}_{j}\!^{i}$. The metric tensor is uniquely determined by
 \begin{equation}
 g_{ij}=e_{i}\!^{\mu}e_{j}\!^{\nu}{\eta}_{\mu\nu}, 
 \end{equation}
 where the Minkowski metric ${\eta}_{\mu\nu}={\rm diag}(-1, +1, +1, +1)$.

In terms of a set of space-time coordinates ({\it i.e.} a holonomic frame), the affine connection of the Riemann-Cartan geometry can be cast in the form
\begin{equation}
{\Gamma} _{ij} \!^{k} =\bar{\Gamma }_{ij} \!^{k} +\frac{1}{2} \left(F_{ij} \!^{k} +F^{k} \!_{ij} +F^{k} \!_{ji} \right) ,
\end{equation} 
where $F_{ij} \! ^{k} $ is the torsion tensor in the holonomic frame, and $\bar{\Gamma }_{ij} \!^{k}$ is the Levi-Civita connection (the Christoffel symbol)
\begin{equation}
\bar{\Gamma }_{ij} \!^{k} =\frac{1}{2} g^{km} (g_{mj,i} +g_{mi,j} -g_{ij,m} ).
\end{equation}
Accordingly, the affine curvature tensors, Ricci curvature, and scalar curvature can be obtained by 
\begin{eqnarray}
R_{ijk} \!^{l} &=&\partial _{i} \Gamma _{jk} \!^{l} -\partial _{j} \Gamma _{ik} \!^{l} +\Gamma _{im} \!^{l} \Gamma _{jk} \!^{m} -\Gamma _{jm} \!^{l} \Gamma _{ik} \!^{m}, \\
R_{ij} &=&R_{kij} \!^{k}, \\
R&=&g^{ij} R_{ij},
\end{eqnarray}
respectively.

%
%
\subsection{The field equations}
The conventional PGT action assume the form as~\cite{PGTHehl},   
\begin{equation}
\begin{aligned}
W=\int & d^{4} x e[L_{m}(\eta _{\alpha \beta ,} ... ,\psi ,D_{\alpha } \psi ) \\
 & +L_{g}(\kappa _{1} ,\kappa _{2} , ... ,\eta _{\alpha \beta } ,F_{\alpha \beta } \!^{\gamma } ,R_{\alpha \beta \gamma} \!^{\delta } )] ,
\end{aligned}
\end{equation}
where $e=\mathrm{det}(e_{i} \!^{\mu})$, and $\kappa_{1},\kappa_{2}, ... $ are some coupling constants. Apparently, the action is dictated by the geometric gauge field Lagrangian density $eL_{g}(e_{i} \!^{\mu} ,\partial_{j} e_{i} \!^{\mu} ,\Gamma_{i \mu} \!^{\nu} ,\partial_{j} \Gamma_{i \mu} \!^{\nu})=eL_{g}(e_{i} \!^{\mu} ,F_{ij} \!^{\mu} ,R_{ij \mu} \!^{\nu})$, as well as by the minimally coupled source Lagragian density $eL_{m}(e_{i}\!^{\mu},\Gamma_{i \mu} \!^{\nu},\psi,\partial_{i}\psi)=eL_{m}(e_{i}\!^{\mu},\psi,D_{i}\psi)$, in which $\psi$ represents all the matter fields. 
Varying the action $W$ with respect to the geometric gauge field potentials $e_{i} \!^{\alpha}$ and $\Gamma_{i}\!^{\alpha \beta}$ yields two gravitational field equations as
\begin{align}
\frac{\delta eL_{g}}{\delta e_{i} \,^{\alpha } } &=-\frac{\delta eL_{m}}{\delta e_{i} \,^{\alpha } } \equiv e\Sigma _{\alpha } \!^{i}, \\
\frac{\delta eL_{g}}{\delta \Gamma _{i} \! ^{\alpha \beta } } &=-\frac{\delta eL_{m}}{\delta \Gamma _{i} \! ^{\alpha \beta } } \equiv e S_{\alpha \beta } \!^{i} ,
\end{align} 
where the source terms $\Sigma _{\alpha } \!^{i} $ and $S _{\alpha \beta } \!^{i} $ represent the canonical momentum current and the canonical spin current governed by the corresponding energy-momentum and angular momentum conservation laws, respectively. Carrying out variations on the left-hand side of Eqs. (10) and (11), one obtains the so-called the first (or translational) and the second (or rotational) gravitational gauge field equations as, respectively,
%
\begin{align}
D_{j} H_{\alpha } \!^{ij} -\varepsilon _{\alpha } \!^{i} &=e\Sigma _{\alpha } \!^{i} ,\\
D_{j} H_{\alpha \beta } \!^{ij} -\varepsilon _{\alpha \beta } \!^{i} &=e S_{\alpha \beta } \!^{i} ,
\end{align} 
where the translational field momenta are given by
\begin{equation}
H_{\alpha } \!^{ij} = \frac{\partial eL_{g}}{\partial \partial_{j} e_{i} \!^{\alpha}}  =2\frac{\partial eL_{g}}{\partial F_{ji} \!^{\alpha } } ,
\end{equation}
the rotatal field momenta are described by
\begin{equation}
H_{\alpha \beta } \!^{ij} = \frac{\partial eL_{g}}{\partial \partial_{j} \Gamma_{i} \!^{\alpha\beta}} =2\frac{\partial eL_{g}}{\partial R_{ji} \!^{\alpha \beta } } ,
\end{equation}
the momentum current (energy-momentum density) is defined as
\begin{equation}
\varepsilon _{\alpha } \!^{i} = e_{\alpha } \!^{i}  eL_{g}-F_{\alpha j} \!^{\gamma } H_{\gamma } \!^{ji} -R_{\alpha j} \!^{\gamma \delta } H_{\gamma \delta } \!^{ji} ,
\end{equation} 
and the spin current (spin angular momentum density) is characterized by
\begin{equation}
\varepsilon _{\alpha \beta } \!^{i} = H_{[\beta \alpha ]} \!^{i} .
\end{equation}

The general Lagrangian density associated with the scalar curvature, quadratic torsion and curvature developed by Baekler and Hehl~\cite{BnH} can be written as 
\begin{equation}
\begin{aligned}
L_{\rm BH} &= \frac{1}{2\kappa } R +\frac{1}{4} F_{\alpha \beta } \!^{\gamma }  
  (d_{1} F_{\gamma } \!^{\alpha \beta } +d_{2} F_{\gamma } \!^{\beta \alpha } +d_{3} \delta _{\gamma } ^{\beta } F_{\mu } \!^{\alpha \mu }  )  \\ 
  &-\frac{1}{4\xi } R_{\alpha \beta \gamma \delta } [R^{\alpha \beta \gamma \delta } +f_{1} R ^{\alpha \gamma \beta \delta } +f_{2} R^{\gamma \delta \alpha \beta } \\ &+f_{3} \eta ^{\alpha \delta } R^{\beta \gamma }
  +f_{4} \eta ^{\alpha \delta } R^{\gamma \beta } +f_{5} \eta ^{\alpha \delta } \eta ^{\beta \gamma } R] ,
\end{aligned}
\end{equation}
where $\kappa,\xi,f_{A},d_{A}$ correspond to various dimensionless coupling constants. Adopting this form as the gravitational Lagrangian density $L_g$, the first equation (12) becomes
\begin{equation}
\begin{aligned}
&\frac{1}{\kappa} (F_{\alpha } \!^{i} -\frac{1}{2} Fe^{i} \!_{\alpha } ) \\
&+\frac{1}{e} D_{j} (ed_{1} F^{ji} \!_{\alpha } +ed_{2} F_{\alpha } \!^{\left[ij\right]} +ed_{3} e^{\left[i\right. } \!_{\alpha } F^{\left. j\right]\gamma } \!_{\gamma } ) \\
&+F_{\alpha j} \!^{\gamma } (d_{1} F^{ij} \!_{\gamma } +d_{2} F_{\gamma } \!^{\left[ji\right]} +d_{3} e^{\left[j\right. } \!_{\gamma } F^{\left. i\right]\mu } \!_{\mu } ) \\
&-\frac{1}{4} e^{i} \!_{\alpha } F_{\mu \nu } \!^{\lambda } (d_{1} F^{\mu \nu } \!_{\lambda } +d_{2} F_{\lambda } \!^{\nu \mu } +d_{3} \delta ^{\nu } _{\lambda } F^{\mu \gamma } \!_{\gamma } ) \\
&+\frac{1}{\xi } R_{\alpha jt} \!^{\gamma \delta } (R^{ji} \!_{\gamma \delta } +f_{1} R^{[j} \!_{\gamma } {} ^{i]} {} \!_{\delta } +f_{2} R_{\gamma \delta } \!^{ji} \\
&+f_{3} e^{[j} \!_{[\delta } R^{i]} \!_{\gamma ]} +f_{4} e^{[j} \!_{[\delta } R_{\gamma ]} \!^{i]} +f_{5} Re^{j} \!_{[\delta } e^{i} \!_{\gamma ]} ) \\
&-\frac{1}{4\xi } e^{i} _{\alpha } R_{\mu \nu \gamma \delta } [R^{\nu \mu \gamma \delta } +f_{1} R^{\nu \gamma \mu \delta } +f_{2} R^{\gamma \delta \nu \mu } \\
&+f_{3} \eta ^{\nu \delta } R^{\mu \gamma } +f_{4} \eta ^{\nu \delta } R^{\gamma \mu } +f_{5} \eta ^{\nu \delta } \eta ^{\gamma \mu } R] =\Sigma _{\alpha }\!^{i},
\end{aligned}
\end{equation}
and the second equation (13) can be explicitly recast as
\begin{equation}
\begin{aligned}
&\frac{\xi }{2\kappa } (F _{\alpha \beta } \!^{i} +2e^{i} \!_{[\alpha }  F_{\beta ]k} \!^{k} ) \\
&+\frac{1}{e} D_{j} (eR^{ij} \!_{\alpha \beta } +ef_{1} R^{[i} \!_{[\alpha } {} ^{j]} {} \!_{\beta ]} +ef_{2} R_{\alpha \beta } \!^{ij} \\ 
&+ef_{3} e^{[i} \!_{[\beta } R^{j]} \!_{\alpha ]} +ef_{4} e^{[i} \!_{[\beta } R_{\alpha ]} \!^{j]} +ef_{5} Re^{i} \!_{[\beta } e^{j} \!_{\alpha ]} ) \\
&+\xi (d_{1} F^{i} \!_{[\beta \alpha ]} -\frac{1}{2} d_{3} e^{i} \!_{[\alpha } F_{\beta ]} \!^{\mu } {} \!_{\mu } +\frac{3}{4} d_{2} F_{[\alpha \beta } \!^{i]} \\
&+\frac{1}{4} d_{2} F_{\alpha \beta } \!^{i} )  =\xi S _{\alpha \beta } \!^{i}.
\end{aligned}
\end{equation}
However complicated they may be, we shall use these field equations directly to probe the dynamics of the universe.
%

%
%
\section{Teleparallel Poincar\'{e} Torsion}
We now apply the PGT formalism to the large scale universe. Under spherical symmetry and spatial reflection invariance, there are only six non-null independent components in the torsion tensor $F_{\alpha \beta \gamma }$ (antisymmetric in the first couple of subscripts $\alpha, \beta$) in spherical coordinates~\cite{BnH}. The cosmological principle requires that these quantities only depend upon time. Accordingly, we assume these non-vanishing torsional components to take the forms as   
\begin{equation}
\begin{aligned}
&F_{010} =-F_{100} =-f(t) , \\
&F_{011} =-F_{101} =-h(t) , \\
&F_{122} =F_{133} =-F_{212} =-F_{313} =g(t) ,\\ 
&F_{022} =F_{033} =-F_{202} =-F_{303} =-\chi(t) ,
\end{aligned}
\end{equation}                         
where $f,h,g,\chi$ are some torsion functions to be determined.

The large scale homogeneity and isotropy can be modeled by the Friedmann-Robertson-Walker(FRW) metric in spherical coordinates as
\begin{equation}
ds^2 =-dt^2 + a(t) ^2 \left[ \frac{dr^2}{1-k r^2} + r^2(d \theta ^2 + \sin ^{2}\theta  d \phi ^2) \right] ,
\end{equation}
where $a(t)$ represents the cosmic scale factor, and the curvature index $k=0, +1, -1$ correspond to, respectively,  a flat, a closed, and an open universe. According to Eq. (3), the FRW metric is uniquely determined by the following set of non-trivial tetrads:
\begin{equation}
\begin{aligned} 
e_{t} \! ^{0} &=1,&
e_{r} \! ^{1} =\frac{a}{\sqrt{1-k r^2}}, \\
e_{\theta } \! ^{2} &=ar,&
e_{\phi } \! ^{3} =ar\sin \theta ,\\
\end{aligned} 
\end{equation}
which associate with dual forms as, respectively,
\begin{equation}
\begin{aligned}
e_{~0}^{t} &=1, 
&e_{~1}^{r} =\frac{\sqrt{1-k r^2}}{a} , \\
e_{~2}^{\theta } &=\frac{1}{ar} , 
&e_{~3}^{\phi } =\frac{1}{ar\sin \theta } .
\end{aligned}
\end{equation}
Once the set of tetrads is chosen, the local Lorentz symmetry~\cite{barrow1,barrow2} is no longer an issue, and one has the freedom to work out the teleparallel gravity theory in either a holonomic frame (spacetime coordinates) or an anholonomic frame (tetrads) since both frames should deliver identical results. This is a merit worthy mentioning because one can perform consistency checks correspondingly. Here, we present our investigation in terms of spacetime coordinates. 

The torsion $F_{ij} \! ^{k} $ in a holonomic frame can be translated from the torsion $F_{\alpha \beta}\!^ {\gamma }$ in an anholonomic frame through the relation of $F_{ij} \!^{k} =e_{i} \!^{\alpha } e_{j} \!^{\beta } e_{\gamma } \!^{k} F_{\alpha \beta } \!^{\gamma }$.  Accordingly, the six non-zero components of torsion assume the form as
\begin{equation}
\begin{aligned}
F_{tr} \!^{t} &=-F_{rt} \!^{t} =\frac{af}{\sqrt{1-k r^2}} ,
&F_{tr} \!^{r} =-F_{rt} \!^{r}=-h ,\\
F_{r\phi } \!^{\phi } &=-F_{\phi r} \!^{\phi }=\frac{ag}{\sqrt{1-k r^2}} ,
&F_{t\phi } \!^{\phi } =-F_{\phi t} \!^{\phi }=-\chi ,\\
F_{r\theta } \!^{\theta } &=-F_{\theta r} \!^{\theta }=\frac{ag}{\sqrt{1-k r^2}} ,
&F_{t\theta } \!^{\theta } =-F_{\theta t} \!^{\theta }=-\chi.
\end{aligned}
\end{equation}
In the other hand, all non-trivial components of affine curvature tensors, Ricci curvatures, and the scalar curvature can be obtained by means of Eqs. (6)-(8). Since the torsion and the curvature are convoluted in accordance with the affine connection [see Eqs. (1) and (2)], one is able to determine the exact form of torsion tensor by imposing the teleparallelism condition, i.e. insisting the absolute parallel property when transporting a given vector along a curve. 

It is straightforward but quite tedious to work out all the teleparallelism constraints. They consist of six independent equations from the null curvature tensors, 
\begin{align}
&\dot{h}+\frac{\dot{a}}{a} h+\frac{\ddot{a}}{a} =0, \\
&gh+\frac{\dot{a}}{a} g+\frac{\chi-h}{ar} \sqrt{1-kr^2}=0 ,\\
&\dot{\chi}+\frac{\dot{a}}{a} \chi+fg+\frac{\ddot{a}}{a} -\frac{f}{ar}\sqrt{1-kr^2}=0,  \\
&\frac{k}{a^2} + h\chi+\frac{\dot{a}}{a} \chi+\frac{\dot{a}}{a}h+\frac{\dot{a}^2}{a^2} +\frac{g}{a r}\sqrt{1-kr^2}=0 ,  \\
&\frac{k}{a^2} + \chi^{2}+ 2\frac{\dot{a}}{a}\chi-g^{2} +\frac{\dot{a}^{2}}{a^2} + 2\frac{g}{ar} \sqrt{1-kr^2}=0 ,\\
&f\chi+\dot{g}+\frac{\dot{a}}{a}g+\frac{\dot{a}}{a}f=0, 
\end{align}
three equations from the vanished Ricci curvatures,
\begin{equation}
2\dot{\chi}+2\frac{\dot{a}}{a} \chi+\dot{h}+\frac{\dot{a}}{a} h+2fg+3\frac{\ddot{a}}{a} -\frac{2f}{ar}\sqrt{1-kr^2}=0, \\
\end{equation}
\begin{equation}
\frac{2k}{a^2}+2h\chi+2\frac{\dot{a}}{a} \chi+\dot{h}+3\frac{\dot{a}}{a} h+\frac{\ddot{a}}{a} +2\left(\frac{\dot{a}}{a} \right)^{2} +\frac{2g}{ar}\sqrt{1-kr^2}=0, \\
\end{equation}
\begin{equation}
\begin{aligned}
&\frac{2k}{a^2}+\dot{\chi}+\chi^{2} +h\chi+4\frac{\dot{a}}{a} \chi+\frac{\dot{a}}{a} h-g^{2} +fg+\frac{\ddot{a}}{a} \\
&+2\left(\frac{\dot{a}}{a} \right)^{2} +\frac{3g-f}{ar}\sqrt{1-kr^2}=0 ,
\end{aligned}
\end{equation}
and one equation obtained by the null scalar curvature,
\begin{equation}
\begin{aligned}
&\frac{6k}{a^2} +  4\dot{\chi}+2\chi^{2} +4h\chi+12\chi\frac{\dot{a}}{a} +2\dot{h}+6h\frac{\dot{a}}{a} -2g^{2} \\
&+4fg +6\frac{\ddot{a}}{a} +6\left(\frac{\dot{a}}{a} \right)^{2} +\frac{8g-4f}{ar}\sqrt{1-kr^2} =0.
\end{aligned}
\end{equation}
The only consistent result fulfilling these curvatureless Eqs. (26)-(35) is that $g=f=0$, and $h=\chi$ with two constraints
\begin{align}
\chi^2 + 2\frac{\dot{a}}{a} +\frac{\dot{a}^2}{a^2} +\frac{k}{a^2}=0,\\
\dot{\chi}+\frac{\dot{a}}{a} \chi+\frac{\ddot{a}}{a} =0.
\end{align} 
As a consequence, the torsion functions are 
\begin{equation}
g=f=0 ,~~~h=\chi=\frac{\sqrt{-k}}{a}-\frac{\dot{a}}{a} ,
\end{equation}
and we obtain the non-null torsion components as~\footnote{Equation (38) is a short handed notation for consistency checks for $k\neq 0$ models. The complex value for $k=+1$ case naturally arises due to the comoving coordinate $r$ used in the metric Eq. (22). It would disappear if we employed another comoving coordinate $x$ defined by integrating over $dx=(1-kr^2)^{-1/2}dr$. }
\begin{equation}
\begin{aligned}
F_{tr} \!^{r} =-F_{rt} \!^{r} &=F_{t\theta } \!^{\theta } =-F_{\theta t} \!^{\theta }  \\
&=F_{t\phi } \!^{\phi } =-F_{\phi t} \!^{\phi } =\frac{\dot{a}}{a}-\frac{\sqrt{-k}}{a}.
\end{aligned}
\end{equation}
Evidently, the form of the Poincar\'{e} torsion is strongly constrained by imposing the teleparallelism condition.

%
\section{Teleparallel Poincar\'{e} cosmology}

Imposing the teleparallism condition to PGT not only strongly constrain the form of the Poincar\'{e} torsion, but also greatly simplifies the field equations (19) and (20) to retain merely quadratic terms in torsion as a result of revoking corresponding curvature terms. In this section, we would like to explore the torsion effect on the cosmological dynamics and extend our scheme to including a spinning fluid as a source driving the cosmic acceleration. To match the current observations, it is sufficient to consider only the case of a flat universe with $k=0$. 
%

%
\subsection{TPGT as a generalized TEGR}
It is natural to consider the intrinsic spin of elementary particles as one of the physical sources for torsion. However, it is usually assumed that the averaged spin density of matter is tiny and randomly oriented in macroscopic regimes~\citep{GRspin,SNY}. The hypothesis amounts to neglect the spin angular momentum tensor $S_{ijk}$ in the second field equation. This is exactly the situation that the standard GR concentrates on.

For a spatially flat universe, $k=0$ and Eq. (39) gives rise to the non-trivial components of the Poincar\'{e} torsion as
\begin{equation}
F_{tr} \!^{r} =-F_{rt} \!^{r} =F_{t\theta } \!^{\theta } =-F_{\theta t} \!^{\theta } =F_{t\phi } \!^{\phi } =-F_{\phi t} \!^{\phi } 
=\frac{\dot{a}}{a} .
\end{equation}
Substituting Eq. (40) into the second field equation (20), we obtain the nontrivial components of the spin angular  momentum tensor as
\begin{eqnarray}
a\dot a \zeta &=& S_{trr} = -S_{rtr} ,\\
a\dot a\zeta r^2 &=& S_{t \theta \theta} = -S_{\theta t \theta} ,\\
a\dot a \zeta r^2 \sin^2{\theta} &=& S_{t \phi \phi} = -S_{\phi t \phi} ,
\end{eqnarray}
where 
\begin{equation}
\zeta=\frac{1}{2} d_{1} + \frac{1}{4} d_{2} +\frac{3}{4} d_{3} -\frac{1}{\kappa }.
\end{equation}
If one is to neglect the spin angular momentum, i.e. requiring that $S_{ijk}=0$, then $\zeta=0$ as long as the universe keeps developing. Thus, the dimensionless coupling constants $d_1, d_2$ and $d_3$ must satisfy the following constraint
\begin{equation}
\frac{1}{2} d_{1} + \frac{1}{4} d_{2} +\frac{3}{4} d_{3}  =\frac{1}{\kappa }.
\end{equation}

In the other hand, substituting Eq. (40) into Eq. (19), the first field equation gives rise to 
\begin{align}
 {3\left( \frac{1}{2} d_{1} + \frac{1}{4} d_{2} +\frac{3}{4} d_{3}\right)}\left(\frac{\dot{a}}{a} \right)^{2} &=\Sigma_{tt}, \\
- {6\left(\frac{1}{2} d_{1} + \frac{1}{4} d_{2} +\frac{3}{4} d_{3}\right)}\left(\frac{\ddot{a}}{a} \right) &=\Sigma,
\end{align}
where $\Sigma_{tt}$ is the time-time component of the energy-momentum tensor of the source, and $\Sigma$ represents the trace of it. Assuming a perfect fluid as the matter source under consideration, then its energy-momentum tensor would take the form of
\begin{equation}
\Sigma_{ij}=(\rho + p)u_{i}u_{j} + p g_{ij},
\end{equation}
where $u_{i}$ is the four-velocity, $\rho$ denotes the energy density, $p=w\rho$ is the pressure, and $w$ represents the equation of state. As a consequence, we have $\Sigma_{tt}=\rho$ and the trace $\Sigma=\rho+3p=(1+3w)\rho$. Along with the constraint Eq. (45), Eqs. (46) and (47) become
%
\begin{align}
H^2 &= \left(\frac{\dot{a}}{a} \right)^{2} =\frac{\kappa }{3} \rho , \\
\frac{\ddot{a}}{a} &=-\frac{\kappa }{6} \left(1 +3w\right)\rho,
\end{align}
where $H\equiv\dot a/a$ is the Hubble parameter characterizing the rate of the cosmic expansion.
One immediately finds that, when ignoring the spin angular momentum, the TPGT recovers the Friedmann cosmology provided that
\begin{equation}
\kappa=8\pi G
\end{equation}
with $G$ being the Newtonian gravitational constant. In this sense, TPGT can be considered as a generalized version of TEGR.

%
%
\subsection{Torsion effects on the cosmic expansion}

In general, a torsion scalar $\mathcal{T}$ can be defined as
\begin{equation}
\mathcal{T} ^2 \equiv F_{\alpha \beta } \!^{\gamma }  
  (F_{\gamma } \!^{\alpha \beta } + F_{\gamma } \!^{\beta \alpha } + \delta _{\gamma } ^{\beta } F_{\mu } \!^{\alpha \mu }  ).
\end{equation}
Subsequently, Eq. (40) indicates that, in a spatially flat universe, the torsion scalar reads
\begin{equation}
\mathcal{T} = 3\frac{\dot{a}}{a} .
\end{equation}
If we take a time derivative of the torsion scalar, then
\begin{equation}
\mathcal{\dot{T}}=3 \left[\frac{\ddot{a}}{a}-\left(\frac{\dot{a}}{a}\right)^2 \right].
\end{equation}
Comparing Eqs. (49)-(50) with Eqs. (53)-(54), a simple relation holds between the equation of state, the torsion scalar,  and its time derivative as
\begin{equation}
\mathcal{\dot{T}}=-{1\over 2}(1+w)\mathcal{T}^2.
\end{equation}
The current observation using the baryon acoustic oscillations (BAO) and CMB data indicates that the equation of state for the expanding universe is always negative but close to $-1$~\cite{planck13}. 
Accordingly, Eq. (55) with $w=-1$ shows that $\mathcal{\dot T} = 0$, i.e. the torsion scalar $\mathcal{T}$ is truly a constant no matter how the universe changes. As a consequence, the universe exhibits a de Sitter expansion with
a scale factor evolving as
\begin{equation}
a(t)=a_{0} \exp\left({\mathcal{T}\over 3} t\right),
\end{equation}
where $a_{0}$ is an integration constant. Hence, the torsion scalar $\mathcal{T}$ is playing the role of the cosmological constant, which is naturally encoded in the teleparallel Poincar\'{e} cosmology. 

%
%
\subsection{Universe containing a spin fluid}

Having shown that the framework of TPGT can be consistently blended in GR, we now extend the theory to include the spin angular momentum as the source of torsion. With the help of a four-velocity  $u_i$, the macroscopic spin tensor of a curvatureless universe can be characterized as
\begin{equation}
S_{ijk} = u_{i}S_{jl} S_{k} \!^{l} - u_{j} S_{jl} S_{k}\! ^ {l},
\end{equation}
where $S_{ij}$ is the spin density tensor in a holonomic frame. Because the intrinsic spin is space-like, the Frenkel condition~\cite{spinfluid2}
\begin{equation}
S_{jk}u^k=0
\end{equation}
is automatically satisfied. The direct products of two spin density tensors in Eq. (57) allows us to contract them after transforming the couple to an anholonomic frame. We then obtain the square of spin density $S^2$ as 
\begin{equation}
S^2 = S_{\alpha\beta} S^{\alpha\beta}.
\end{equation}
As a consequence, the macroscopic spin tensor becomes
\begin{equation}
S_{ijk}=\left( u_{i} g_{jk}-u_{j} g_{ik} \right)S^2.
\end{equation}
Employing this form in Eqs. (41)-(43), the nontrivial components of the spin angular momentum tensor generated by the second equation are governed by
\begin{align}
a \dot{a} \zeta &= -S^2 a^2,\\
a \dot{a} r^2 \zeta &= -S^2 a^2 r^2,\\
a \dot{a} r^2 \sin{\theta}^2 \zeta &= -S^2 a^2 r^2 \sin{\theta}^2 ,
\end{align}
which unanimously lead to a constraint to the square of spin density,
\begin{equation}
S^2=-\frac{\dot{a}}{a} \zeta .
\end{equation}
Substituting Eq. (64) into Eqs. (46)-(47), the Friedmann equation becomes
\begin{equation}
\left(\frac{\dot{a}}{a} \right)^{2} =\frac{\kappa }{3} \rho + \kappa \frac{\dot{a}}{a}  S^2,
\end{equation}
and the acceleration equation is given by
\begin{equation}
\frac{\ddot{a}}{a} =-\frac{\kappa }{6}\left(1+3w \right)\left(\rho + 3 S^2 \frac{\dot{a}}{a} \right).
\end{equation}
In order to unravel the role played by intrinsic spins from convoluting with the gravitational effect, we assume that the spin density tensor $S_{\alpha\beta}$ can be characterized by the torsion scalar $\mathcal{T}$ as
\begin{equation}
S_{\alpha\beta}=\frac{1}{\sqrt{2 \mathcal{T}}}s_{\alpha\beta},
\end{equation}
where $s_{\alpha\beta}$ denotes the reduced spin density such that $s^2 = \frac{1}{2} s_{\alpha\beta} s^{\alpha\beta}$. Accordingly, Eqs. (65) and (66) become
%
\begin{eqnarray}
\left(\frac{\dot{a}}{a} \right)^{2} &=&\frac{\kappa }{3}\rho_{\rm eff},\\
\frac{\ddot{a}}{a} &=&-\frac{\kappa }{6} (1+3w)\rho_{\rm eff},
\end{eqnarray}
with the effective energy density of the spin fluid described by
\begin{equation}
\rho_{\rm eff}=\rho+s^2\equiv\rho+\rho_s.
\end{equation}
That is, we recover the standard Friedmann formulation for a homogeneous and isotropic universe. 
We note that, according to Eq. (70), the square of spin density has positive contribution to the total energy density of the universe, contrary to the Weyssenhoff spin fluid in the framework of the Einstein-Carton theory~\cite{spinfluid4}. In addition, a dust spin fluid model with $w=0$ is incapable of explaining the late-time cosmic acceleration. 

Now let us assume Eqs. (68) and (69) be describing the evolution of the expanding universe driven by 
the energy content of the vacuum. Then, under the condition that the vacuum energy $p=-\rho \rightarrow 0$ ($w=-1$),
the spin energy would serve as an alternative of dark energy responsible for cosmic acceleration. Therefore, the cosmological constant $\Lambda$ is no more than a uniform spin energy density in the context of the spinning TPGT.

%
%
%

\section{Conclusion} 

In this work we investigated the cosmological effects of the TPGT. Adopting the specific torsion form constrained by the teleparallelism condition, the formulation of the teleparallel Poincar\'{e} cosmology is completely equivalent to that of the Friedmann cosmology. As a consequence, the naturally built-in torsion scalar $\mathcal{T}$ plays the role of the cosmological constant $\Lambda$ and the universe exhibits a de Sitter expansion provided that the equation of state $w=-1$. 

When extending the teleparallel framework to include a spinning fluid as the source of the Poincar\'{e} torsion, the standard Friedmann prescription to a homogeneous and isotropic universe is recovered, as long as the macroscopic spin tensor satisfies Eq. (60) with a spin density governed by Eq. (67). Contributing positively to the effective energy density of the universe, such a dust spinning fluid with $w=0$ acts just like a normal matter. Under the circumstances that the vacuum energy with $w=-1$ vanishes, however, the vacuum spin may be considered as the source driving the late-time cosmic acceleration. People attempt to solve the cosmological constant problem, namely, why the dark energy that we observe is so much smaller than any known energy scales or otherwise it is zero as protected, if there is any, by a symmetry. The results here may open another window for us to understand the cosmological constant problem. 
While the vacuum energy is thought to be originated from zero-point energy fluctuations, the spin energy 
may be associated with spin fluctuations of the vacuum that should be very different from zero-point energies.
Understanding the microscopic spin structure of the vacuum is a very interesting subject and its effect on 
the cosmological scales as alluded here should be further studied.

%
\begin{acknowledgments}
The authors are grateful to Wai Bong Yeung and Jim Nester for helpful suggestions and discussions. This work was supported in part by the National Science Council, Taiwan, ROC under the Grant No. NSC101-2112-M-001-010-MY3, and by the Office of Research and Development, National Taiwan Normal University, Taiwan, ROC under the Grant No. 102A05. 
\end{acknowledgments}

\nocite{*}
\bibliography{tpgt1}

\begin{thebibliography}{37}%
\makeatletter
\providecommand \@ifxundefined [1]{%
 \@ifx{#1\undefined}
}%
\providecommand \@ifnum [1]{%
 \ifnum #1\expandafter \@firstoftwo
 \else \expandafter \@secondoftwo
 \fi
}%
\providecommand \@ifx [1]{%
 \ifx #1\expandafter \@firstoftwo
 \else \expandafter \@secondoftwo
 \fi
}%
\providecommand \natexlab [1]{#1}%
\providecommand \enquote  [1]{``#1''}%
\providecommand \bibnamefont  [1]{#1}%
\providecommand \bibfnamefont [1]{#1}%
\providecommand \citenamefont [1]{#1}%
\providecommand \href@noop [0]{\@secondoftwo}%
\providecommand \href [0]{\begingroup \@sanitize@url \@href}%
\providecommand \@href[1]{\@@startlink{#1}\@@href}%
\providecommand \@@href[1]{\endgroup#1\@@endlink}%
\providecommand \@sanitize@url [0]{\catcode `\\12\catcode `\$12\catcode
  `\&12\catcode `\#12\catcode `\^12\catcode `\_12\catcode `\%12\relax}%
\providecommand \@@startlink[1]{}%
\providecommand \@@endlink[0]{}%
\providecommand \url  [0]{\begingroup\@sanitize@url \@url }%
\providecommand \@url [1]{\endgroup\@href {#1}{\urlprefix }}%
\providecommand \urlprefix  [0]{URL }%
\providecommand \Eprint [0]{\href }%
\providecommand \doibase [0]{http://dx.doi.org/}%
\providecommand \selectlanguage [0]{\@gobble}%
\providecommand \bibinfo  [0]{\@secondoftwo}%
\providecommand \bibfield  [0]{\@secondoftwo}%
\providecommand \translation [1]{[#1]}%
\providecommand \BibitemOpen [0]{}%
\providecommand \bibitemStop [0]{}%
\providecommand \bibitemNoStop [0]{.\EOS\space}%
\providecommand \EOS [0]{\spacefactor3000\relax}%
\providecommand \BibitemShut  [1]{\csname bibitem#1\endcsname}%
\let\auto@bib@innerbib\@empty
\bibitem [{\citenamefont {Riess}\ \emph {et~al.}(1998)\citenamefont {Riess}
  \emph {et~al.}}]{Ia1}%
  \BibitemOpen
  \bibfield  {author} {\bibinfo {author} {\bibfnamefont {A.~G.}\ \bibnamefont
  {Riess}} \emph {et~al.},\ }\href@noop {} {\bibfield  {journal} {\bibinfo
  {journal} {Astron. J.}\ }\textbf {\bibinfo {volume} {116}},\ \bibinfo {pages}
  {1009} (\bibinfo {year} {1998})}\BibitemShut {NoStop}%
\bibitem [{\citenamefont {Perlmutter}\ \emph {et~al.}(1999)\citenamefont
  {Perlmutter} \emph {et~al.}}]{Ia2}%
  \BibitemOpen
  \bibfield  {author} {\bibinfo {author} {\bibfnamefont {S.}~\bibnamefont
  {Perlmutter}} \emph {et~al.},\ }\href@noop {} {\bibfield  {journal} {\bibinfo
   {journal} {Astrophys. J.}\ }\textbf {\bibinfo {volume} {517}},\ \bibinfo
  {pages} {565} (\bibinfo {year} {1999})}\BibitemShut {NoStop}%
\bibitem [{\citenamefont {Komatsu}\ \emph {et~al.}(2011)\citenamefont {Komatsu}
  \emph {et~al.}}]{wmap7}%
  \BibitemOpen
  \bibfield  {author} {\bibinfo {author} {\bibfnamefont {E.}~\bibnamefont
  {Komatsu}} \emph {et~al.},\ }\href@noop {} {\bibfield  {journal} {\bibinfo
  {journal} {Astrophys. J. Suppl.}\ }\textbf {\bibinfo {volume} {192}},\
  \bibinfo {pages} {18} (\bibinfo {year} {2011})}\BibitemShut {NoStop}%
\bibitem [{\citenamefont {Tegmark}\ \emph {et~al.}(2004)\citenamefont {Tegmark}
  \emph {et~al.}}]{sdss}%
  \BibitemOpen
  \bibfield  {author} {\bibinfo {author} {\bibfnamefont {M.}~\bibnamefont
  {Tegmark}} \emph {et~al.},\ }\href@noop {} {\bibfield  {journal} {\bibinfo
  {journal} {Phys. Rev. D}\ }\textbf {\bibinfo {volume} {69}},\ \bibinfo
  {pages} {103501} (\bibinfo {year} {2004})}\BibitemShut {NoStop}%
\bibitem [{\citenamefont {Weinberger}(1989)}]{cc}%
  \BibitemOpen
  \bibfield  {author} {\bibinfo {author} {\bibfnamefont {S.}~\bibnamefont
  {Weinberger}},\ }\href@noop {} {\bibfield  {journal} {\bibinfo  {journal}
  {Rev. Mod. Phys.}\ }\textbf {\bibinfo {volume} {61}},\ \bibinfo {pages} {1}
  (\bibinfo {year} {1989})}\BibitemShut {NoStop}%
\bibitem [{\citenamefont {Amendola}\ \emph {et~al.}(2006)\citenamefont
  {Amendola}, \citenamefont {Quartin}, \citenamefont {Tsujikawa},\ and\
  \citenamefont {Waga}}]{coin}%
  \BibitemOpen
  \bibfield  {author} {\bibinfo {author} {\bibfnamefont {L.}~\bibnamefont
  {Amendola}}, \bibinfo {author} {\bibfnamefont {M.}~\bibnamefont {Quartin}},
  \bibinfo {author} {\bibfnamefont {S.}~\bibnamefont {Tsujikawa}}, \ and\
  \bibinfo {author} {\bibfnamefont {I.}~\bibnamefont {Waga}},\ }\href@noop {}
  {\bibfield  {journal} {\bibinfo  {journal} {Phys. Rev. D}\ }\textbf {\bibinfo
  {volume} {74}},\ \bibinfo {pages} {023525} (\bibinfo {year}
  {2006})}\BibitemShut {NoStop}%
\bibitem [{\citenamefont {Bengochea}\ and\ \citenamefont
  {Ferraro}(2009)}]{ft1}%
  \BibitemOpen
  \bibfield  {author} {\bibinfo {author} {\bibfnamefont {G.}~\bibnamefont
  {Bengochea}}\ and\ \bibinfo {author} {\bibfnamefont {R.}~\bibnamefont
  {Ferraro}},\ }\href@noop {} {\bibfield  {journal} {\bibinfo  {journal} {Phys.
  Rev. D}\ }\textbf {\bibinfo {volume} {79}},\ \bibinfo {pages} {124019}
  (\bibinfo {year} {2009})}\BibitemShut {NoStop}%
\bibitem [{\citenamefont {Linder}(2010)}]{ft2}%
  \BibitemOpen
  \bibfield  {author} {\bibinfo {author} {\bibfnamefont {E.~V.}\ \bibnamefont
  {Linder}},\ }\href@noop {} {\bibfield  {journal} {\bibinfo  {journal} {Phys.
  Rev. D}\ }\textbf {\bibinfo {volume} {81}},\ \bibinfo {pages} {127301}
  (\bibinfo {year} {2010})}\BibitemShut {NoStop}%
\bibitem [{\citenamefont {Capozzielo}\ \emph {et~al.}(2011)\citenamefont
  {Capozzielo}, \citenamefont {Cardone}, \citenamefont {Farajollahi},\ and\
  \citenamefont {Ravanpak}}]{ft3}%
  \BibitemOpen
  \bibfield  {author} {\bibinfo {author} {\bibfnamefont {S.}~\bibnamefont
  {Capozzielo}}, \bibinfo {author} {\bibfnamefont {V.~F.}\ \bibnamefont
  {Cardone}}, \bibinfo {author} {\bibfnamefont {H.}~\bibnamefont
  {Farajollahi}}, \ and\ \bibinfo {author} {\bibfnamefont {A.}~\bibnamefont
  {Ravanpak}},\ }\href@noop {} {\bibfield  {journal} {\bibinfo  {journal}
  {Phys. Rev. D}\ }\textbf {\bibinfo {volume} {84}},\ \bibinfo {pages} {043527}
  (\bibinfo {year} {2011})}\BibitemShut {NoStop}%
\bibitem [{\citenamefont {Bamba}\ \emph {et~al.}(2012)\citenamefont {Bamba},
  \citenamefont {Myrzakulov}, \citenamefont {Nojiti},\ and\ \citenamefont
  {Ofinydon}}]{ft4}%
  \BibitemOpen
  \bibfield  {author} {\bibinfo {author} {\bibfnamefont {K.}~\bibnamefont
  {Bamba}}, \bibinfo {author} {\bibfnamefont {R.}~\bibnamefont {Myrzakulov}},
  \bibinfo {author} {\bibfnamefont {S.}~\bibnamefont {Nojiti}}, \ and\ \bibinfo
  {author} {\bibfnamefont {S.~D.}\ \bibnamefont {Ofinydon}},\ }\href@noop {}
  {\bibfield  {journal} {\bibinfo  {journal} {Phys. Rev. D}\ }\textbf {\bibinfo
  {volume} {85}},\ \bibinfo {pages} {104036} (\bibinfo {year}
  {2012})}\BibitemShut {NoStop}%
\bibitem [{\citenamefont {Geng}\ \emph {et~al.}(2011)\citenamefont {Geng},
  \citenamefont {Lee}, \citenamefont {Saridakis},\ and\ \citenamefont
  {Wu}}]{ft5}%
  \BibitemOpen
  \bibfield  {author} {\bibinfo {author} {\bibfnamefont {C.-Q.}\ \bibnamefont
  {Geng}}, \bibinfo {author} {\bibfnamefont {C.-C.}\ \bibnamefont {Lee}},
  \bibinfo {author} {\bibfnamefont {E.~N.}\ \bibnamefont {Saridakis}}, \ and\
  \bibinfo {author} {\bibfnamefont {Y.-P.}\ \bibnamefont {Wu}},\ }\href@noop {}
  {\bibfield  {journal} {\bibinfo  {journal} {Phys. Lett. B}\ }\textbf
  {\bibinfo {volume} {704}},\ \bibinfo {pages} {384} (\bibinfo {year}
  {2011})}\BibitemShut {NoStop}%
\bibitem [{\citenamefont {de~Andrade}\ \emph {et~al.}()\citenamefont
  {de~Andrade}, \citenamefont {Gullien},\ and\ \citenamefont
  {Pereira}}]{TGoverview}%
  \BibitemOpen
  \bibfield  {author} {\bibinfo {author} {\bibfnamefont {V.~C.}\ \bibnamefont
  {de~Andrade}}, \bibinfo {author} {\bibfnamefont {L.~C.~T.}\ \bibnamefont
  {Gullien}}, \ and\ \bibinfo {author} {\bibfnamefont {J.~G.}\ \bibnamefont
  {Pereira}},\ }\href@noop {} {\enquote {\bibinfo {title} {Teleparallel
  gravity: An overview},}\ }\bibinfo {howpublished} {talk given at the IX
  Marcel Grossmann Meeting, Rome, Italy, July, 2000},\ \bibinfo {note}
  {arXiv:gr-qc/0011087}\BibitemShut {NoStop}%
\bibitem [{\citenamefont {Aldrovandi}\ and\ \citenamefont
  {Pereira}()}]{introductionTG}%
  \BibitemOpen
  \bibfield  {author} {\bibinfo {author} {\bibfnamefont {R.}~\bibnamefont
  {Aldrovandi}}\ and\ \bibinfo {author} {\bibfnamefont {J.}~\bibnamefont
  {Pereira}},\ }\href@noop {} {\enquote {\bibinfo {title} {An introduction to
  teleparallel gravity},}\ }\bibinfo {howpublished}
  {http://www.ift.unesp.br/users/jpereira/tele.pdf}\BibitemShut {NoStop}%
\bibitem [{\citenamefont {Maluf}(2013)}]{tegrev}%
  \BibitemOpen
  \bibfield  {author} {\bibinfo {author} {\bibfnamefont {J.~W.}\ \bibnamefont
  {Maluf}},\ }\href@noop {} {\bibfield  {journal} {\bibinfo  {journal} {Annalen
  der Physik}\ }\textbf {\bibinfo {volume} {525}},\ \bibinfo {pages} {339}
  (\bibinfo {year} {2013})}\BibitemShut {NoStop}%
\bibitem [{\citenamefont {Sauer}(2006)}]{einstein}%
  \BibitemOpen
  \bibfield  {author} {\bibinfo {author} {\bibfnamefont {T.}~\bibnamefont
  {Sauer}},\ }\href@noop {} {\bibfield  {journal} {\bibinfo  {journal}
  {Historia Math.}\ }\textbf {\bibinfo {volume} {32}},\ \bibinfo {pages} {339}
  (\bibinfo {year} {2006})},\ \bibinfo {note} {arXiv:
  physics/0405142}\BibitemShut {NoStop}%
\bibitem [{\citenamefont {de~Andrade}\ and\ \citenamefont
  {Pereira}(1997)}]{torsionforce}%
  \BibitemOpen
  \bibfield  {author} {\bibinfo {author} {\bibfnamefont {V.~C.}\ \bibnamefont
  {de~Andrade}}\ and\ \bibinfo {author} {\bibfnamefont {J.~G.}\ \bibnamefont
  {Pereira}},\ }\href@noop {} {\bibfield  {journal} {\bibinfo  {journal} {Phys.
  Rev. D}\ }\textbf {\bibinfo {volume} {56}},\ \bibinfo {pages} {4689}
  (\bibinfo {year} {1997})}\BibitemShut {NoStop}%
\bibitem [{\citenamefont {Arcos}\ and\ \citenamefont
  {Pereira}(2004)}]{TGreapprisal}%
  \BibitemOpen
  \bibfield  {author} {\bibinfo {author} {\bibfnamefont {H.~I.}\ \bibnamefont
  {Arcos}}\ and\ \bibinfo {author} {\bibfnamefont {J.~G.}\ \bibnamefont
  {Pereira}},\ }\href@noop {} {\bibfield  {journal} {\bibinfo  {journal} {Int.
  J. Mod. Phys. D}\ }\textbf {\bibinfo {volume} {13}},\ \bibinfo {pages} {2193}
  (\bibinfo {year} {2004})}\BibitemShut {NoStop}%
\bibitem [{\citenamefont {Sotiriou}\ and\ \citenamefont
  {Faraoni}(2010)}]{frrev}%
  \BibitemOpen
  \bibfield  {author} {\bibinfo {author} {\bibfnamefont {T.~P.}\ \bibnamefont
  {Sotiriou}}\ and\ \bibinfo {author} {\bibfnamefont {V.}~\bibnamefont
  {Faraoni}},\ }\href@noop {} {\bibfield  {journal} {\bibinfo  {journal} {Rev.
  Mod. Phys.}\ }\textbf {\bibinfo {volume} {82}},\ \bibinfo {pages} {451}
  (\bibinfo {year} {2010})}\BibitemShut {NoStop}%
\bibitem [{\citenamefont {Ferraro}\ and\ \citenamefont
  {Fiorini}(2007)}]{FFinf}%
  \BibitemOpen
  \bibfield  {author} {\bibinfo {author} {\bibfnamefont {R.}~\bibnamefont
  {Ferraro}}\ and\ \bibinfo {author} {\bibfnamefont {F.}~\bibnamefont
  {Fiorini}},\ }\href@noop {} {\bibfield  {journal} {\bibinfo  {journal} {Phys.
  Rev. D}\ }\textbf {\bibinfo {volume} {75}},\ \bibinfo {pages} {084031}
  (\bibinfo {year} {2007})}\BibitemShut {NoStop}%
\bibitem [{\citenamefont {Li}\ \emph {et~al.}(2011)\citenamefont {Li},
  \citenamefont {Sotiriou},\ and\ \citenamefont {Barrow}}]{barrow1}%
  \BibitemOpen
  \bibfield  {author} {\bibinfo {author} {\bibfnamefont {B.}~\bibnamefont
  {Li}}, \bibinfo {author} {\bibfnamefont {T.~P.}\ \bibnamefont {Sotiriou}}, \
  and\ \bibinfo {author} {\bibfnamefont {D.}~\bibnamefont {Barrow}},\
  }\href@noop {} {\bibfield  {journal} {\bibinfo  {journal} {Phys. Rev. D}\
  }\textbf {\bibinfo {volume} {83}},\ \bibinfo {pages} {064035} (\bibinfo
  {year} {2011})}\BibitemShut {NoStop}%
\bibitem [{\citenamefont {Sotiriou}\ \emph {et~al.}(2011)\citenamefont
  {Sotiriou}, \citenamefont {Li},\ and\ \citenamefont {Barrow}}]{barrow2}%
  \BibitemOpen
  \bibfield  {author} {\bibinfo {author} {\bibfnamefont {T.~P.}\ \bibnamefont
  {Sotiriou}}, \bibinfo {author} {\bibfnamefont {B.}~\bibnamefont {Li}}, \ and\
  \bibinfo {author} {\bibfnamefont {D.}~\bibnamefont {Barrow}},\ }\href@noop {}
  {\bibfield  {journal} {\bibinfo  {journal} {Phys. Rev. D}\ }\textbf {\bibinfo
  {volume} {83}},\ \bibinfo {pages} {104030} (\bibinfo {year}
  {2011})}\BibitemShut {NoStop}%
\bibitem [{\citenamefont {Fiorini}\ and\ \citenamefont
  {Ferraro}(2011)}]{FFcos}%
  \BibitemOpen
  \bibfield  {author} {\bibinfo {author} {\bibfnamefont {F.}~\bibnamefont
  {Fiorini}}\ and\ \bibinfo {author} {\bibfnamefont {R.}~\bibnamefont
  {Ferraro}},\ }\href@noop {} {\bibfield  {journal} {\bibinfo  {journal} {Int.
  J. Mod. Phys. Conf. Ser.}\ }\textbf {\bibinfo {volume} {3}},\ \bibinfo
  {pages} {227} (\bibinfo {year} {2011})}\BibitemShut {NoStop}%
\bibitem [{\citenamefont {Hehl}(1980)}]{PGTHehl}%
  \BibitemOpen
  \bibfield  {author} {\bibinfo {author} {\bibfnamefont {F.~W.}\ \bibnamefont
  {Hehl}},\ }in\ \href@noop {} {\emph {\bibinfo {booktitle} {Cosmology and
  Gravitation on Spin, Torsion and Supergravity}}},\ \bibinfo {editor} {edited
  by\ \bibinfo {editor} {\bibfnamefont {P.~G.}\ \bibnamefont {Bergamann}}\ and\
  \bibinfo {editor} {\bibfnamefont {V.}~\bibnamefont {de~Sabatta}}}\ (\bibinfo
  {publisher} {Plenmun, New York},\ \bibinfo {year} {1980})\ p.~\bibinfo
  {pages} {5}\BibitemShut {NoStop}%
\bibitem [{\citenamefont {Baekler}\ and\ \citenamefont {Hehl}(1983)}]{BnH}%
  \BibitemOpen
  \bibfield  {author} {\bibinfo {author} {\bibfnamefont {P.}~\bibnamefont
  {Baekler}}\ and\ \bibinfo {author} {\bibfnamefont {F.~W.}\ \bibnamefont
  {Hehl}},\ }in\ \href@noop {} {\emph {\bibinfo {booktitle} {Lecture Notes in
  Physics: Gauge Theory and Gravitation}}},\ \bibinfo {editor} {edited by\
  \bibinfo {editor} {\bibfnamefont {K.}~\bibnamefont {Kikkawa}}, \bibinfo
  {editor} {\bibfnamefont {N.}~\bibnamefont {Nakanishi}}, \ and\ \bibinfo
  {editor} {\bibfnamefont {H.}~\bibnamefont {Nariai}}}\ (\bibinfo  {publisher}
  {Springer-Verlag Berlin Heidelberg},\ \bibinfo {year} {1983})\ p.~\bibinfo
  {pages} {1}\BibitemShut {NoStop}%
\bibitem [{\citenamefont {Blagojevic}(2003)}]{3lecturesPGT}%
  \BibitemOpen
  \bibfield  {author} {\bibinfo {author} {\bibfnamefont {M.}~\bibnamefont
  {Blagojevic}},\ }\href@noop {} {\bibfield  {journal} {\bibinfo  {journal}
  {SFIN A}\ }\textbf {\bibinfo {volume} {1}},\ \bibinfo {pages} {147} (\bibinfo
  {year} {2003})},\ \bibinfo {note} {arXiv: gr-qc/0302040}\BibitemShut
  {NoStop}%
\bibitem [{\citenamefont {Goenner}\ and\ \citenamefont {{F.
  M\"{u}ller-Hoissen}}(1984)}]{goenner}%
  \BibitemOpen
  \bibfield  {author} {\bibinfo {author} {\bibfnamefont {H.}~\bibnamefont
  {Goenner}}\ and\ \bibinfo {author} {\bibnamefont {{F. M\"{u}ller-Hoissen}}},\
  }\href@noop {} {\bibfield  {journal} {\bibinfo  {journal} {Class. Quant.
  Grav.}\ }\textbf {\bibinfo {volume} {1}},\ \bibinfo {pages} {651} (\bibinfo
  {year} {1984})}\BibitemShut {NoStop}%
\bibitem [{\citenamefont {Yo}\ and\ \citenamefont
  {Nester}(2007)}]{OscillatingUniverse}%
  \BibitemOpen
  \bibfield  {author} {\bibinfo {author} {\bibfnamefont {H.-J.}\ \bibnamefont
  {Yo}}\ and\ \bibinfo {author} {\bibfnamefont {J.~M.}\ \bibnamefont
  {Nester}},\ }\href@noop {} {\bibfield  {journal} {\bibinfo  {journal} {Mod.
  Phys. Lett. A}\ }\textbf {\bibinfo {volume} {22}},\ \bibinfo {pages} {2057}
  (\bibinfo {year} {2007})}\BibitemShut {NoStop}%
\bibitem [{\citenamefont {Shie}\ \emph {et~al.}(2008)\citenamefont {Shie},
  \citenamefont {Nester},\ and\ \citenamefont {Yo}}]{SNY}%
  \BibitemOpen
  \bibfield  {author} {\bibinfo {author} {\bibfnamefont {K.-F.}\ \bibnamefont
  {Shie}}, \bibinfo {author} {\bibfnamefont {J.~M.}\ \bibnamefont {Nester}}, \
  and\ \bibinfo {author} {\bibfnamefont {H.-J.}\ \bibnamefont {Yo}},\
  }\href@noop {} {\bibfield  {journal} {\bibinfo  {journal} {Phys. Rev. D}\
  }\textbf {\bibinfo {volume} {78}},\ \bibinfo {pages} {023522} (\bibinfo
  {year} {2008})}\BibitemShut {NoStop}%
\bibitem [{\citenamefont {Ao}\ and\ \citenamefont {Li}(2012)}]{lxz2012}%
  \BibitemOpen
  \bibfield  {author} {\bibinfo {author} {\bibfnamefont {X.-C.}\ \bibnamefont
  {Ao}}\ and\ \bibinfo {author} {\bibfnamefont {X.-Z.}\ \bibnamefont {Li}},\
  }\href@noop {} {\bibfield  {journal} {\bibinfo  {journal} {JCAP}\ }\textbf
  {\bibinfo {volume} {02}},\ \bibinfo {pages} {003} (\bibinfo {year}
  {2012})}\BibitemShut {NoStop}%
\bibitem [{\citenamefont {Tseng}\ \emph {et~al.}(2012)\citenamefont {Tseng},
  \citenamefont {Lee},\ and\ \citenamefont {Geng}}]{gengpgt}%
  \BibitemOpen
  \bibfield  {author} {\bibinfo {author} {\bibfnamefont {H.-H.}\ \bibnamefont
  {Tseng}}, \bibinfo {author} {\bibfnamefont {C.-C.}\ \bibnamefont {Lee}}, \
  and\ \bibinfo {author} {\bibfnamefont {C.-Q.}\ \bibnamefont {Geng}},\
  }\href@noop {} {\bibfield  {journal} {\bibinfo  {journal} {JCAP}\ }\textbf
  {\bibinfo {volume} {11}},\ \bibinfo {pages} {013} (\bibinfo {year}
  {2012})}\BibitemShut {NoStop}%
\bibitem [{\citenamefont {Hehl}\ \emph {et~al.}(1976)\citenamefont {Hehl},
  \citenamefont {von~der Heyde}, \citenamefont {Kerlick},\ and\ \citenamefont
  {Nester}}]{GRspin}%
  \BibitemOpen
  \bibfield  {author} {\bibinfo {author} {\bibfnamefont {F.~W.}\ \bibnamefont
  {Hehl}}, \bibinfo {author} {\bibfnamefont {P.}~\bibnamefont {von~der Heyde}},
  \bibinfo {author} {\bibfnamefont {G.~D.}\ \bibnamefont {Kerlick}}, \ and\
  \bibinfo {author} {\bibfnamefont {J.~M.}\ \bibnamefont {Nester}},\
  }\href@noop {} {\bibfield  {journal} {\bibinfo  {journal} {Rev. Mod. Phys.}\
  }\textbf {\bibinfo {volume} {48}},\ \bibinfo {pages} {393} (\bibinfo {year}
  {1976})}\BibitemShut {NoStop}%
\bibitem [{\citenamefont {Smalley}\ and\ \citenamefont
  {Krisch}(1994)}]{spinfluid3}%
  \BibitemOpen
  \bibfield  {author} {\bibinfo {author} {\bibfnamefont {L.~L.}\ \bibnamefont
  {Smalley}}\ and\ \bibinfo {author} {\bibfnamefont {J.~P.}\ \bibnamefont
  {Krisch}},\ }\href@noop {} {\bibfield  {journal} {\bibinfo  {journal} {Class.
  Quant. Grav.}\ }\textbf {\bibinfo {volume} {11}},\ \bibinfo {pages} {2375}
  (\bibinfo {year} {1994})}\BibitemShut {NoStop}%
\bibitem [{\citenamefont {Hehl}\ \emph {et~al.}(1974)\citenamefont {Hehl},
  \citenamefont {von~der Heyde},\ and\ \citenamefont {Kerlick}}]{spinfluid1}%
  \BibitemOpen
  \bibfield  {author} {\bibinfo {author} {\bibfnamefont {F.~W.}\ \bibnamefont
  {Hehl}}, \bibinfo {author} {\bibfnamefont {P.}~\bibnamefont {von~der Heyde}},
  \ and\ \bibinfo {author} {\bibfnamefont {G.~D.}\ \bibnamefont {Kerlick}},\
  }\href@noop {} {\bibfield  {journal} {\bibinfo  {journal} {Phys. Rev. D}\
  }\textbf {\bibinfo {volume} {10}},\ \bibinfo {pages} {1066} (\bibinfo {year}
  {1974})}\BibitemShut {NoStop}%
\bibitem [{\citenamefont {Obukhov}\ and\ \citenamefont
  {Korotky}(1987)}]{spinfluid2}%
  \BibitemOpen
  \bibfield  {author} {\bibinfo {author} {\bibfnamefont {Y.~N.}\ \bibnamefont
  {Obukhov}}\ and\ \bibinfo {author} {\bibfnamefont {V.~A.}\ \bibnamefont
  {Korotky}},\ }\href@noop {} {\bibfield  {journal} {\bibinfo  {journal}
  {Class. Quant. Grav.}\ }\textbf {\bibinfo {volume} {4}},\ \bibinfo {pages}
  {1633} (\bibinfo {year} {1987})}\BibitemShut {NoStop}%
\bibitem [{\citenamefont {Szyd{\l}owski}\ and\ \citenamefont
  {Krawiec}(2004)}]{spinfluid4}%
  \BibitemOpen
  \bibfield  {author} {\bibinfo {author} {\bibfnamefont {M.}~\bibnamefont
  {Szyd{\l}owski}}\ and\ \bibinfo {author} {\bibfnamefont {A.}~\bibnamefont
  {Krawiec}},\ }\href@noop {} {\bibfield  {journal} {\bibinfo  {journal} {Phys.
  Rev. D}\ }\textbf {\bibinfo {volume} {70}},\ \bibinfo {pages} {043510}
  (\bibinfo {year} {2004})}\BibitemShut {NoStop}%
\bibitem [{Note1()}]{Note1}%
  \BibitemOpen
  \bibinfo {note} {Equation (38) is a short handed notation for consistency
  checks for $k\not =0$ models. The complex value for $k=+1$ case naturally
  arises due to the comoving coordinate $r$ used in the metric Eq. (22). It
  would disappear if we employed another comoving coordinate $x$ defined by
  integrating over $dx=(1-kr^2)^{-1/2}dr$.}\BibitemShut {Stop}%
\bibitem [{\citenamefont {{Planck Collaboration}}()}]{planck13}%
  \BibitemOpen
  \bibfield  {author} {\bibinfo {author} {\bibnamefont {{Planck
  Collaboration}}},\ }\href@noop {} {\ }\bibinfo {note}
  {{arXiv:1303.5076}}\BibitemShut {NoStop}%
\end{thebibliography}%
\end{document}